\documentclass{PoS}

\usepackage{amsmath,amssymb,amsfonts} \usepackage[english]{babel}
\usepackage{cite}

\usepackage{ulem}

\newcommand{\A}{{a}} 
 
\newcommand{\gGF}{g_{\rm GF}^2}
\newcommand{\MSb}{\overline{\textrm{MS}}}

\definecolor{title}{rgb}{0.20,0.20,0.70}
\definecolor{emph}{rgb}{0.18,0.18,0.60}
\definecolor{emph1}{rgb}{0.18,0.18,0.60}
\definecolor{emph2}{rgb}{0.70,0.18,0.18}

\title{Gradient flow running coupling in SU(2) with $N_f=6$ flavors}

\ShortTitle{Gradient flow running coupling in SU(2) with 6 flavors}

\author{\speaker{Viljami Leino}\\ Helsinki Institute of Physics and
		Department of Physics, University of Helsinki\\ E-mail:
		\email{viljami.leino@helsinki.fi}}

\author{Teemu Rantalaiho\\ 
		Helsinki Institute of Physics and Department
		of Physics, University of Helsinki\\ E-mail:
		\email{teemu.rantalaiho@helsinki.fi}}

\author{Kari Rummukainen\\
		Helsinki Institute of Physics and Department
		of Physics, University of Helsinki\\ E-mail:
		\email{kari.rummukainen@helsinki.fi}}

\author{Joni M. Suorsa\\
		Helsinki Institute of Physics and Department
		of Physics, University of Helsinki\\ E-mail:
		\email{joni.suorsa@helsinki.fi}}

\author{Kimmo Tuominen \\
		Helsinki Institute of Physics and Department
		of Physics, University of Helsinki\\ E-mail:
		\email{kimmo.i.tuominen@helsinki.fi}}

\author{Sara Tähtinen \\
		Helsinki Institute of Physics and Department
		of Physics, University of Helsinki\\ E-mail:
		\email{sara.tahtinen@helsinki.fi}}

\abstract{
We present preliminary results of the running of the coupling in
SU(2) gauge theory with 6 massless fundamental representation fermion flavors.
We measure the coupling using the gradient flow method with Schrödinger functional boundary conditions.
The results are consistent with perturbation theory in the weak coupling and
we see an indication of infrared fixed point at strong coupling.
}

\FullConference{34th annual International Symposium on Lattice Field Theory\\
		24-30 July 2016\\
		University of Southampton, UK}

\begin{document}

\section{Introduction} 
Many phenomenologically viable models of beyond standard model physics can be built
on asymptotically free gauge theories where the running of the coupling
approaches a non-trivial infrared fixed point (IRFP) and the long distance physics becomes conformal.
Given a SU($N_c$) gauge theory with $N_f$ massless flavors of Dirac fermions,
we can locate theories with an IRFP by choosing a fermion representation and varying the
number of fermions $N_f$.
The range of values $N_f$ where the theory has an IRFP is called the conformal window.
The upper edge of conformal window can be calculated from perturbation theory
by finding the $N_f$ where the one-loop coefficient of the $\beta$-function vanishes.
Over this region the theory loses its asymptotic freedom. 
However, as the $N_f$ is lowered, the IRFP will shift towards larger couplings
until a spontaneous chiral chiral symmetry breaking occurs and theory becomes QCD-like.
The smallest $N_f$ that still has a IRFP behavior marks the lower edge of the conformal window.
This lower boundary is typically located at strong coupling,
which mandates the use of nonperturbative methods, such as lattice simulations, 
to determine its location.
Over recent years this question has been heavily studied in multiple different models.

In this paper we focus on SU(2) gauge theory with six flavors of fundamental representation massless fermions.
This theory is supposed to be near the lower edge of the conformal window,
which is estimated to be between $N_f\sim6-8$ by different approximations~\cite{Sannino:2004qp,Dietrich:2006cm,Frandsen:2010ej}.
From previous lattice studies the $N_f=4$ and $N_f=8,\,10$ cases are known to be
outside and inside of the conformal window respectively~\cite{Leino:2015bfg,Ohki:2010sr,Karavirta:2011zg}.
Direct searches for the presence or the lack of an IRFP for the six fermion case 
have, however, been inconclusive~\cite{Karavirta:2011zg,Bursa:2010xn,Hayakawa:2013maa,Appelquist:2013pqa}.

We employ the the gradient flow finite volume method~\cite{Luscher:2011bx,Luscher:2010iy},
with Schrödinger functional boundary conditions~\cite{Luscher:1991wu,Fritzsch:2013je} 
to measure the running of the coupling constant.
This allows us to reach vanishing fermion mass and measure the mass anomalous dimension alongside the coupling~\cite{suorsa}.
We run the analysis with multiple discretizations and find a clear indication of IRFP at $\gGF\sim13-15$.

\section{Methods and Results}
In this work we study the SU(2) gauge theory with six massless Dirac fermions
in the fundamental representation. We use the HEX smeared~\cite{Capitani:2006ni}, 
clover improved Wilson fermion action with partially smeared plaquette gauge action
as our lattice formulation:
%\begin{equation}
%\begin{aligned} 
%	S &=(1-c_g)S_G(U) + c_g S_G(V) + S_F(V)\,, \\
%	S_G(U) &= \beta_L \sum_{x;\mu<\nu} 
%   	\left (1 - \frac12 \Tr [U_\mu(x) U_\nu(x+a\hat\mu) 
% 	 U^\dagger_\mu(x+a\hat\nu) U^\dagger_\nu(x) ] \right)\,, \\
%	S_F &= a^4\sum_{\alpha=1}^{N_f} \sum_x \left [
%  	\bar{\psi}_\alpha(x) ( i D + m_0 )
%  	\psi_\alpha(x)
%   	+ a \csw \bar\psi_\alpha(x)\frac{i}{4}\sigma_{\mu\nu}
%  	F_{\mu\nu}(x)\psi_\alpha(x) \right ] \,,
%\end{aligned} 
%\end{equation}
\begin{equation} 
	S = (1-c_g)S_G(U) + c_g S_G(V) + S_F(V) + c_{SW} \delta S_{SW}(V),
\end{equation}%
where $V$ and $U$ are the smeared and unsmeared gauge fields respectively.
The smearing of the standard single plaquette Wilson gauge action $S_G$ is
tuned by the parameter~$c_g$ to remove the unphysical bulk phase transition 
from the region of interest in the parameter space\cite{DeGrand:2011vp}. 
Here we set~$c_g=0.5$. 
The clover Wilson fermion action $S_F$ is non-perturbatively improved to order $\mathcal{O}(\A)$ 
with the tree-level Sheikholeslami-Wohlert coefficient set to ~$c_{SW}\approx1$.

We use the Schrödinger Functional method \cite{Luscher:1991wu}
with  Dirichlet boundary conditions.
On a lattice of size~$L^4$ the gauge fields are set to unity 
and the fermion fields are set to zero at temporal boundaries~$x_0=0,L$:
\begin{equation}
\begin{aligned}
U_k(0,{\bf{x}})&=U_k(L,{\bf{x}})=V_k(0,{\bf{x}})=V_k(L,{\bf{x}})=1 \,,\\
U_\mu(x_0,{\bf{x}}+L\hat{{\bf{k}}})&=U_\mu(x_0,{\bf{x}})\,,\; V_\mu(x_0,{\bf{x}}+L\hat{{\bf{k}}})=V_\mu(x_0,{\bf{x}})\,,\\
 \psi(0,{\bf{x}})&=\psi(L,{\bf{x}})=0\,,\; \psi(x_0,{\bf{x}}+L\hat{{\bf{k}}}) = \psi(x_0,{\bf{x}}) \,
\end{aligned}
\end{equation}%
where $k$ labels one of the spatial directions.
These boundary conditions enable us to both run simulations at vanishing quark mass, and 
measure the mass anomalous dimension alongside the running coupling.
%The fermion masses are tuned to zero at the $24^4$ sized lattice, using the
%PCAC quark mass relation \cite{Luscher:1996vw}. 
%The $24^4$ parameters are then used for all lattice sizes. 

We measure the running of the coupling using the Yang-Mills gradient flow.
This method is set up by introducing a fictitious flow time $t$
and studying the evolution of the flow gauge field $B_\mu(x,t)$ according to flow equation:
%\begin{align*} 
%  \partial_t B_{t,\mu} &= D_{t,\mu} B_{t,\mu\nu}, \\
%  B_{0,\mu} &= A_\mu\\ G_{t,\mu\nu} &= \partial_\mu B_{t,\nu} -
%               \partial_\nu B_{t,\mu} + \left[ B_{t,\mu},B_{t,\nu} \right].
%\end{align*} 
\begin{equation}
  \partial_t B_{\mu} = D_{\nu} G_{\nu\mu} \,,\; 
\end{equation}%
where $G_{\mu\nu}(x;t)$ is the field strength of the flow field $B_{\mu}$ and $D_\mu=\partial_\mu+[B_\mu,\,\cdot\,]$.
The initial condition is defined such that $B_{\mu}(x;t=0) = A_\mu(x)$ 
in terms of the original continuum gauge field $A_\mu$. 
In the lattice formulation the continuum flow field is replaced by
the lattice link variable $U$, which is then evolved using either the tree-level
improved Lüscher-Weisz pure gauge action (LW)\cite{Luscher:1984xn} or the Wilson plaquette gauge action (W).

\begin{figure}[t]
  \includegraphics[width=0.49\textwidth]{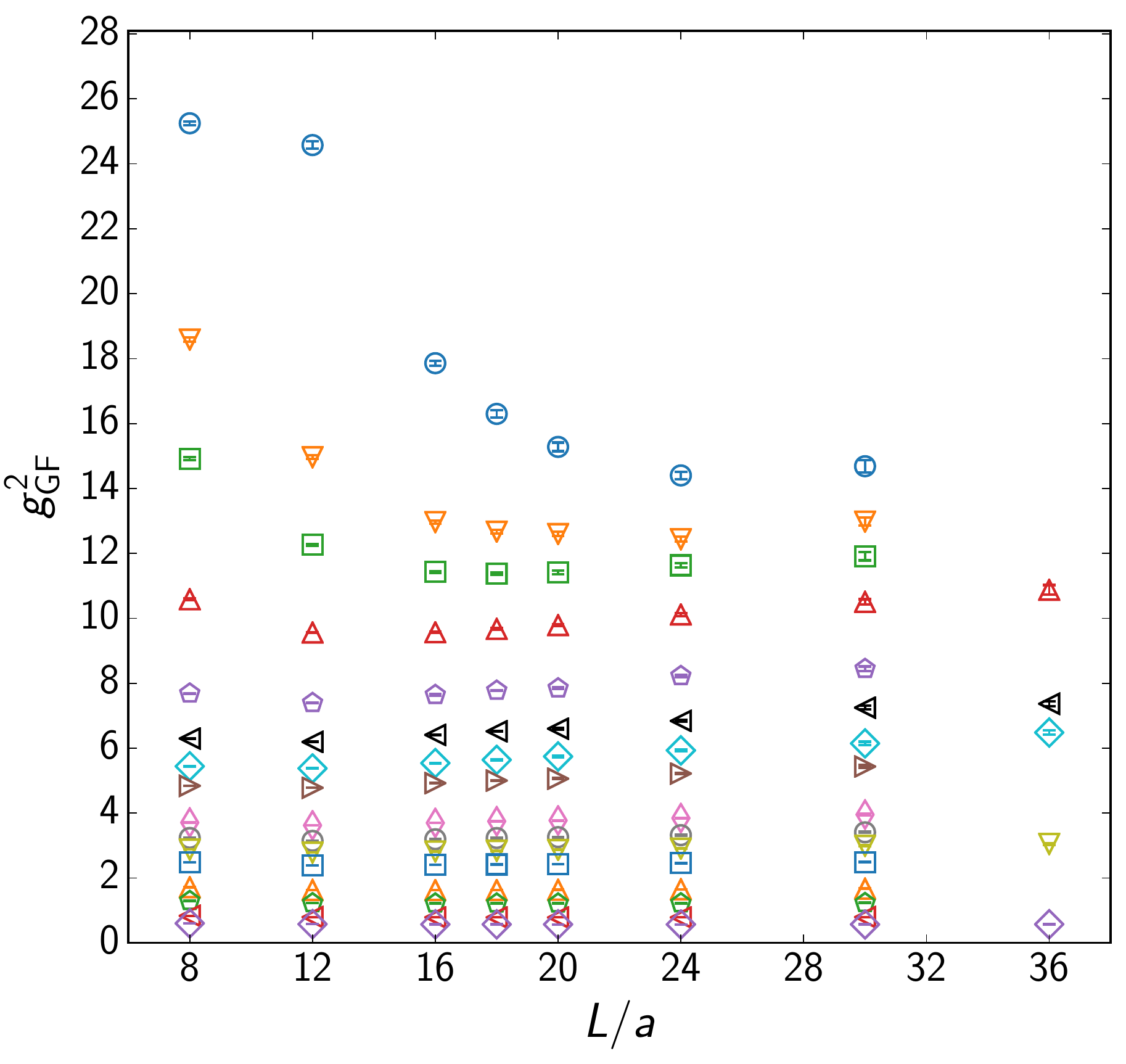}
  \includegraphics[width=0.49\textwidth]{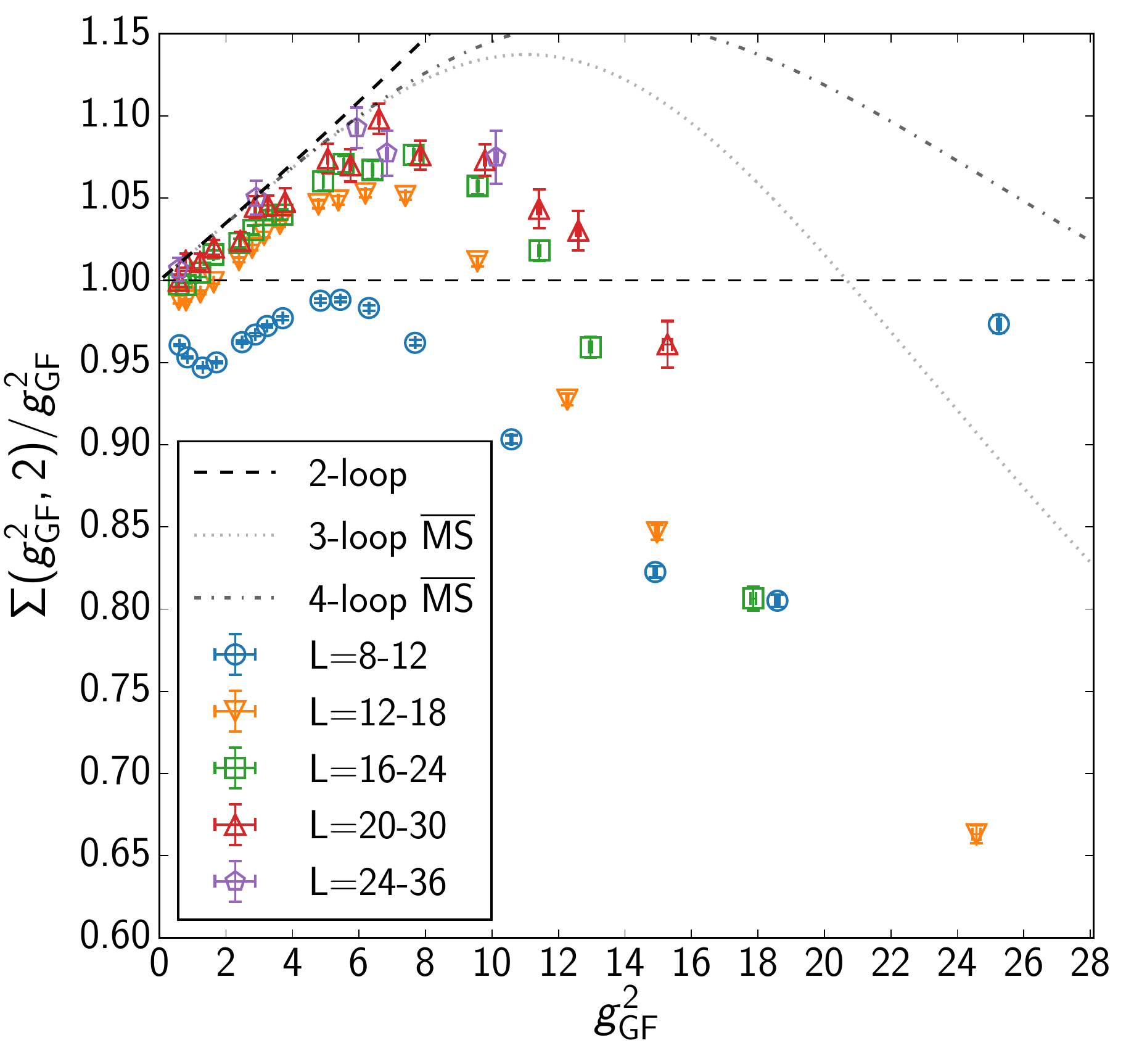} 
  \caption[b]{ 
              The plot on the left shows the gradient flow coupling~\eqref{eq:g2gf}
              measured at each $\beta$ and $L/\A$ at $c=0.4$. 
			  The plot on the right shows the lattice step scaling function~\eqref{lat_step_raw} for these couplings. 
  }
  \label{fig:g2_lat_meas} 
\end{figure}
The flow smooths the gauge field over a radius $\sqrt{8t}$,
removing the UV divergences and automatically renormalizing gauge invariant observables~\cite{Luscher:2011bx}.
Thus we can use evolution of the field strength, to the leading order in perturbation theory in $\overline{\text{MS}}$ scheme,
to define the coupling at scale $\mu=1/\sqrt{8t}$~\cite{Luscher:2010iy}:
\begin{align}
	\label{eq:EE}
    \left<E(t)\right> &= \frac 14 \left<G_{\mu\nu}(t)G_{\mu\nu}(t)\right> = \frac{3(N^2-1)g_0^2}{128\pi^2 t^2}+\mathcal{O}(g_0^4)\,, \\
	\label{eq:g2gf}
	\gGF(\mu) &= \mathcal{N}^{-1}t^2 \left < E(t) \right>\vert_{x_0=L/2\,,\,t=1/8\mu^2}\,,
\end{align}%
where the  normalization factor $\mathcal{N}$ has been calculated in~\cite{Fritzsch:2013je}
for the Schrödinger functional finite size scaling. 
As the translation symmetry is broken by the chosen boundary conditions, 
the coupling $\gGF$ is measured only on the central time slice $x_0=L/2$.
In the lattice formulation we measure the $\left<E(t)\right>$ using both 
symmetric clover and simple plaquette discretizations.

In order to limit the scale into a regime $1/L \ll \mu \ll 1/a$, where \eqref{eq:g2gf} is free of both lattice artifacts and finite volume effects, 
we relate the lattice scale to the renormalization scale by defining a dimensionless parameter $c_t$ 
as described in \cite{Fodor:2012td}:
\begin{equation}
\mu^{-1} = c_tL = \sqrt{8t}.
\end{equation}%
It is suggested in \cite{Fritzsch:2013je} that the SF scheme has reasonably small cutoff effects 
and statistical variance within the range of $c_t=0.3-0.5$.

We choose to do bulk of our analysis with gradient flow evolved with Lüscher-Weisz action,
clover definition of energy density~\eqref{eq:EE}, and $c_t=0.3$. 
Results from these parameters can then be compared with the  other discretizations to 
%gain insight of the systematic errors.
check the reliability of our analysis in the continuum limit.
We run the simulations using lattice sizes 8,12,16,18,20,24,30 and 36 and with bare couplings within the range $g_0^2\in[0.5,8]$.
The measured couplings with the aforementioned parameters are shown in figure~\ref{fig:g2_lat_meas}.
It is clear from the figure that the finite volume effects become substantial on smaller lattices as the
coupling grows larger. Since the measurements on the $L=36$ are incomplete, 
they will not be included in any advanced analysis.

\begin{figure}[t]
  \includegraphics[width=0.49\textwidth]{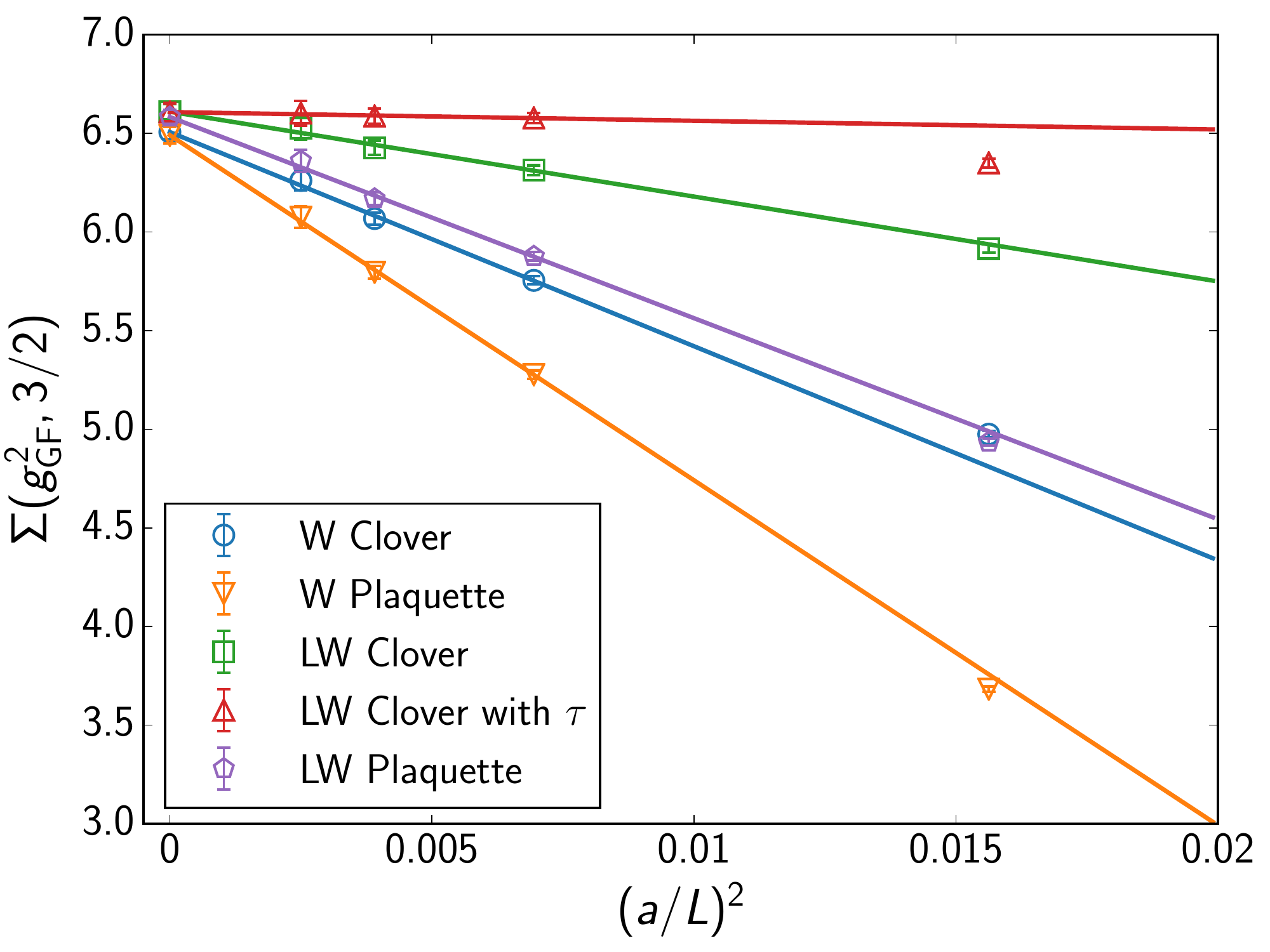}
  \includegraphics[width=0.49\textwidth]{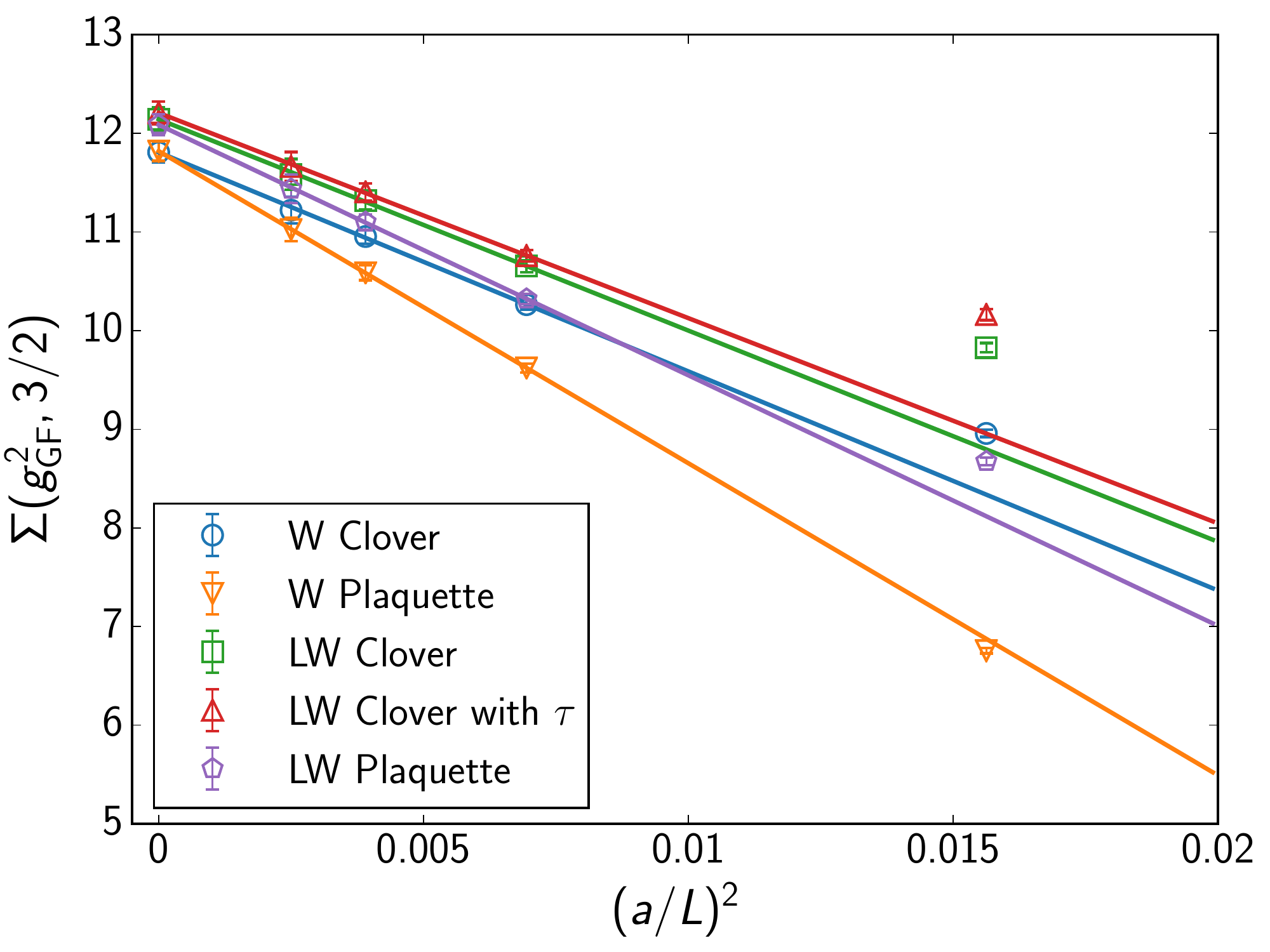} 
  \caption[b]{ The continuum limit~\eqref{lat_step_cont} 
			  with different discretizations and the effect of $\tau_0$-correction.
              Left: $\gGF=6$, Right: $\gGF=11$. The smallest lattice size is not used in the fit.
  }
  \label{fig:cutoff} 
\end{figure}
To quantify the running of the coupling we use the finite lattice spacing step scaling function~\cite{Luscher:1993gh}:
\begin{equation} \label{lat_step_raw} 
    \Sigma(u,L/\A,s) = \left . \gGF(g_0,sL/\A) \right|_{\gGF(g_0,L/\A)=u}\,,
\end{equation}%
which describes the change of the measured coupling 
when the linear size of the system is increased from $L$ to $sL$.
Our data allows us to use either $s=2$ or $s=3/2$. For this paper we have chosen the step size $s=3/2$.
In figure~\ref{fig:g2_lat_meas} we show the scaled step scaling function $\Sigma(u,L/\A,3/2)/u$
calculated for the measured pairs $8-12$, $12-18$, $16-24$, $20-30$ and $24-36$.
The large coupling behavior of the $8-12$ pair deviates significantly from the others
probably due to finite volume effects.

We expect the lowest order discretization effect to be of order $\mathcal{O}(\A^2)$ and 
extrapolate the continuum limit of the step scaling function $\sigma(u)$ with a fit:
\begin{align} \label{lat_step_cont} 
	\Sigma(u,\A/L)  &= \sigma(u) + c(u) (\A/L)^2 \\
	\label{eq:sigco}
	\sigma(u) &= \lim_{\A\rightarrow 0} \Sigma(u,\A/L)\,,
\end{align}%
where we obtain the constant values of couplings at several lattice sizes by interpolating the measured couplings as:
\begin{equation} \label{betafitfun} 
    %\gGF(g_0,\A/L) = g_0^2(1+g_0^2 \sum_{i=0}^m  a_i g_0^{2i}) \,, \;\; m=10\,.
	\frac{\gGF(g_0)}{g_0^4}-\frac{1}{g_0^2}=\sum_{i=0}^m  a_i g_0^{2i} \,, \;\; m=10\,.
\end{equation}%
With this choice of a polynomial function we achieve a combined $\chi^2/$d.o.f of $\sim 1.1$.
We study the robustness of the fit by also running the interpolation with $m=9$ and repeating the analysis.

\begin{figure}[t]
  \includegraphics[width=0.49\textwidth]{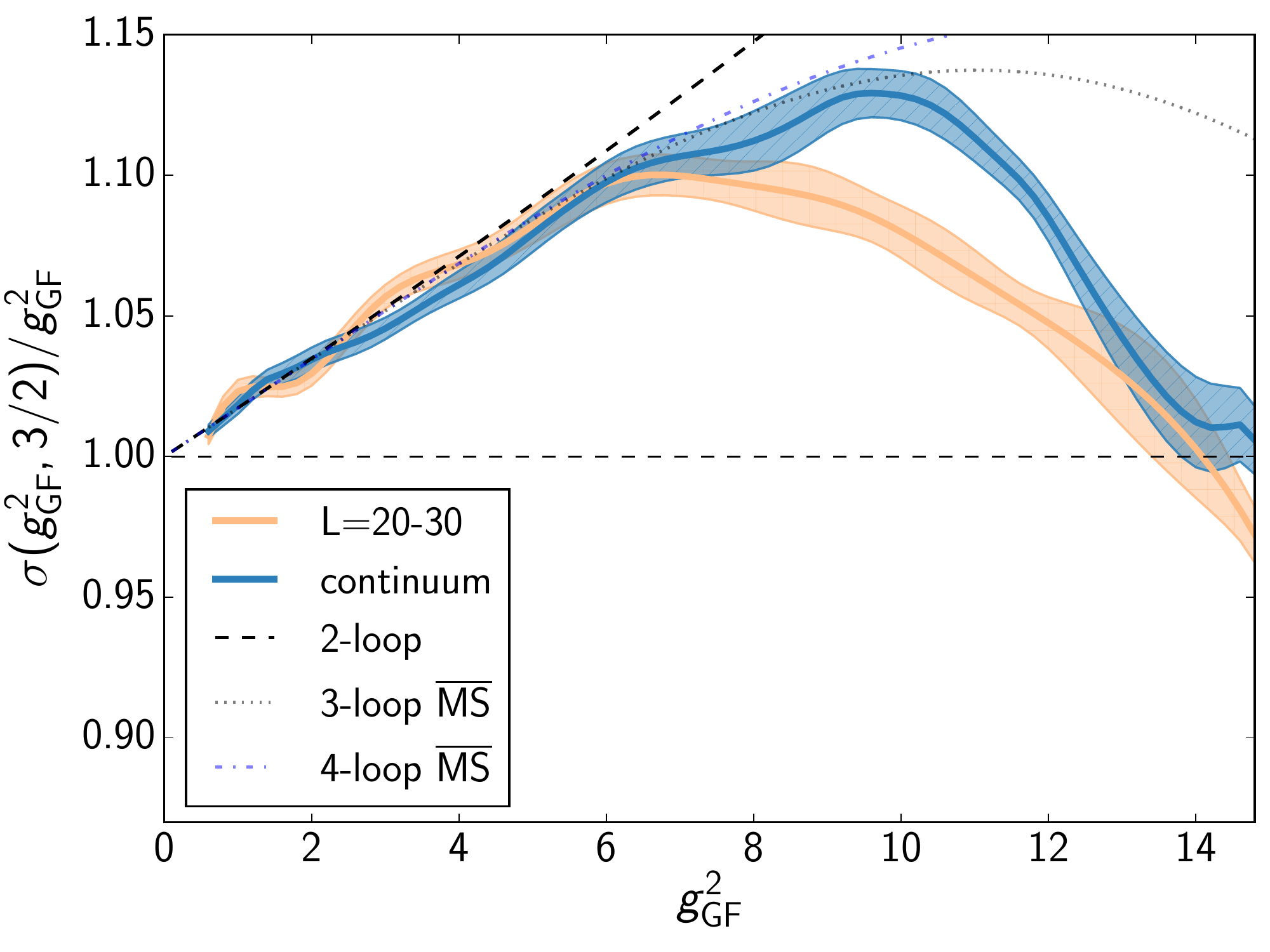}
  \includegraphics[width=0.49\textwidth]{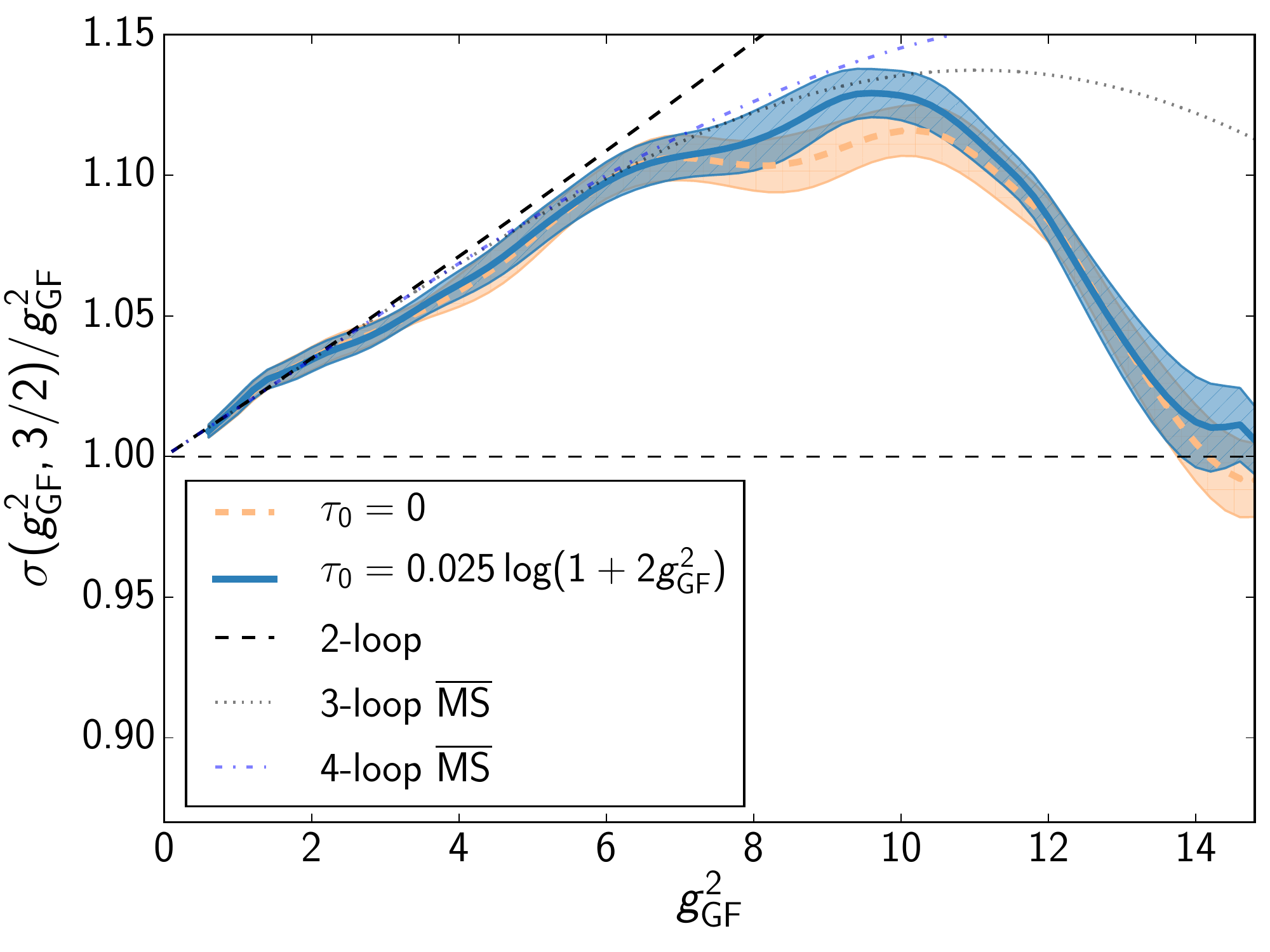}\\
  \includegraphics[width=0.49\textwidth]{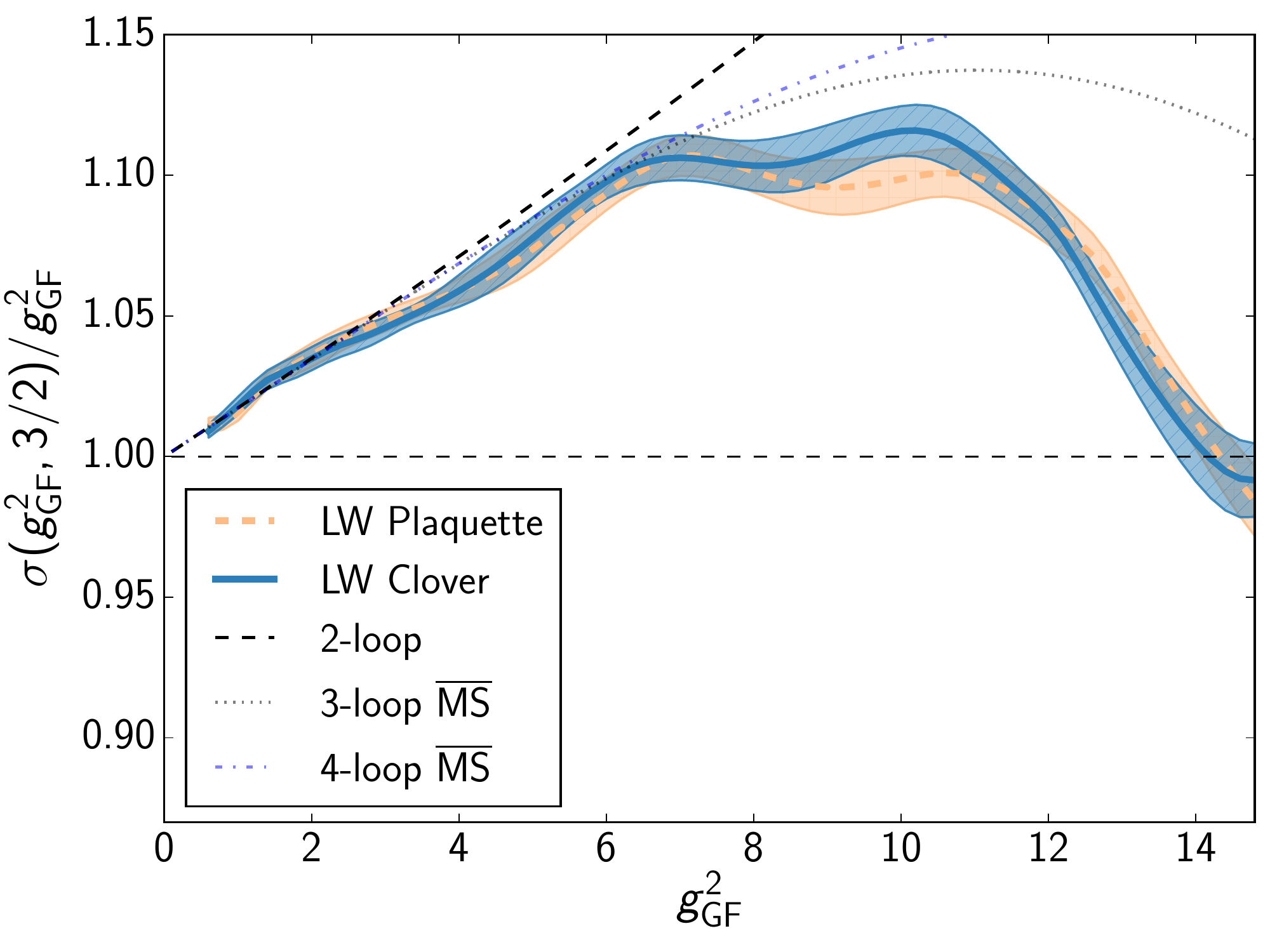}
  \includegraphics[width=0.49\textwidth]{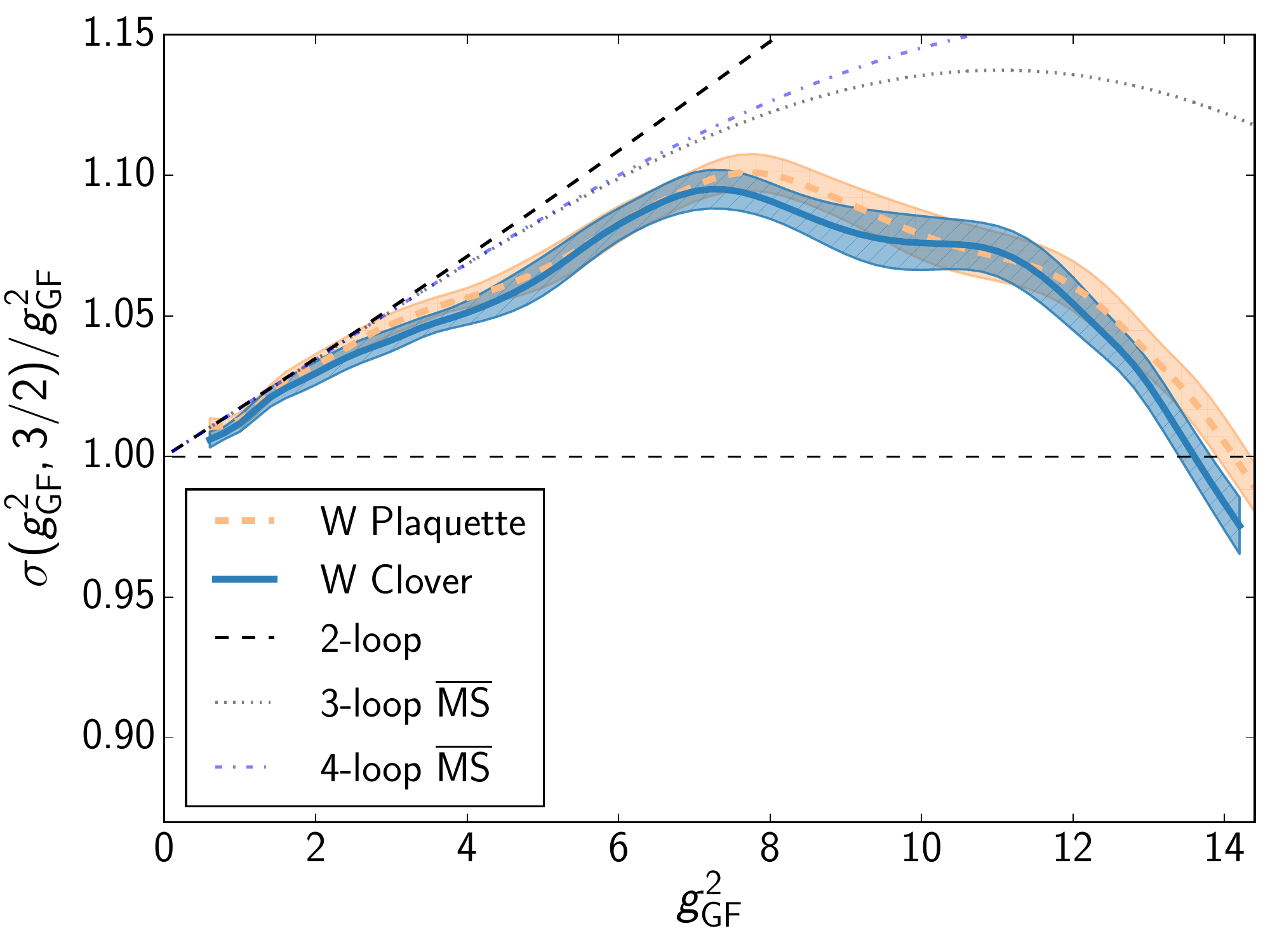}
  \caption[b]{ 
              The scaled step scaling function with continuum extrapolation 
			  calculated and compared with different discretizations.
			  Upper row: LW flow with clover energy with and without $\tau_0$ correction.
			  Lower row: Plaquette and clover $E$'s for LW and W evolved flows.
  }
  \label{fig:contsigma} 
\end{figure}
The continuum limit of the step scaling function~\eqref{lat_step_cont} 
can be used to give an estimate of the cutoff effects.
As the gradient flow coupling is known to produce $\mathcal{O}(\A^2)$ discretization effects
we optimize the gradient flow coupling~\eqref{eq:g2gf} 
to minimize the $\mathcal{O}(\A^2)$ lattice artifacts in the continuum step scaling function
by adding a tunable $\tau_0$ correction to it, as suggested in \cite{Cheng:2014jba}:
\begin{equation} 
\label{eq:taucor}
	\gGF = \frac{t^2}{\mathcal{N}} \langle E(t+\tau_0 a^2) \rangle = 
	\frac{t^2}{\mathcal{N}} \langle E(t) \rangle + \frac{t^2}{\mathcal{N}} 
	\langle \frac{\partial E(t)}{\partial t} \rangle \tau_0 a^2 + \mathcal{O}(a^4)\,.
\end{equation}%
It turns out the precise value of $\tau_0$ has a relatively small effect in the continuum extrapolation,
as long as it is not allowed to grow too large \cite{Hasenfratz:2014rna}. 
For $c_t=0.3$ a constant $\tau_0=0.05$ would suffice for most of the measured couplings, 
but it would be too large and affect the continuum limit for small couplings.
Therefore we have decided to make the $\tau_0$-correction a function of the measured coupling $\gGF$:
\begin{equation}
\tau_0 = 0.025\log(1+2\gGF)\,.
\label{eq:taufunc}
\end{equation}%
The measured coupling is used instead of the bare coupling
in order to have a consistent $\mathcal{O}(\A^2)$ shift in the step scaling analysis\cite{Ramos:2015dla}.
The final $\tau_0$ is then calculated iteratively starting from $\gGF=g_0^2$.

In figure~\ref{fig:cutoff} we show the $\A^2$-dependence of the step scaling function for 
all measured discretizations without any $\tau_0$ correction
% and compare them 
compared to our chosen set of discretizations with the $\tau_0$ correction~\eqref{eq:taufunc} applied.
$\tau_0$ correction removes most of the $\mathcal{O}(\A^2)$ cutoff effects
in the small coupling regime where it was defined to do so.
Generally the more improved discretizations (LW over W, clover over plaquette) seem to have
smaller cutoff effects.
However, we see clear violations on the leading $\mathcal{O}(\A^2)$ scaling on small lattice sizes
and therefore will not use smallest lattice size $L=8$ in our analysis.
Interestingly the Wilson flow with plaquette energy seems to have 
the most consistent $\mathcal{O}(\A^2)$ scaling despite it having the largest cutoff effects.

We present the continuum extrapolations of step scaling function~\eqref{eq:sigco} 
for multiple different discretizations in the figure~\ref{fig:contsigma}.
Similar to lattice step scaling behavior in figure ~\ref{fig:g2_lat_meas},
%Like seen in the lattice step scaling figure~\ref{fig:g2_lat_meas}, 
the continuum step scaling follows the universal two loop perturbative curve closely
up to $\gGF\approx7$ and then diverges towards an IRFP around $\gGF\sim14.5$.
While the 3 and 4-loop $\MSb$ curves are scheme dependent, and cannot be
directly compared, they are shown as a reference.
On the upper left picture where we have the continuum limit with  the chosen set of discretizations and $\tau_0$-correction,
we also show the lattice step scaling of the largest lattice pair $L=20-30$.
From the other pictures we can see all discretizations to mostly agree in the continuum within $1-\sigma$ error bands.

\section{Conclusions} 
We have studied the running coupling in the SU(2) lattice gauge theory 
with 6 fermions in the fundamental representation.
Gradient flow algorithm with Schrödinger functional boundaries 
gives us a clear look to large coupling behavior of this theory.
We see a clear indication of a fixed point around $\gGF\sim13-15$ in the step scaling analysis.
The continuum limit is robust regardless of the discretizations used. 
For added reliability, results with different $c_t$ remain to be calculated.
The results for the mass anomalous dimension are reported in~\cite{suorsa}.
\section{Acknowledgments}
This work is supported by the Academy of Finland grants 267842, 134018 and 267286, 
T.R. and S.T. are funded by the Magnus Ehrnrooth foundation and J.M.S. by the Jenny and Antti Wihuri foundation.
The simulations were performed at the Finnish IT Center for Science (CSC) in Espoo, Finland.
Parts of the simulation program have been derived from the MILC lattice simulation program~\cite{MILC}.

\end{document}